\documentclass[12pt]{iopart}

\usepackage{iopams}
\usepackage[utf8]{inputenc}
\usepackage[english]{babel}

\usepackage{etoolbox}

\expandafter\let\csname equation*\endcsname\relax
\expandafter\let\csname endequation*\endcsname\relax 

\usepackage{amsfonts}
\usepackage{amssymb}
\usepackage{amsmath}
\usepackage{amsthm}
\usepackage{color}
\usepackage{mathrsfs}
\usepackage{relsize}
\usepackage{bm}
\usepackage{hyperref}
\usepackage{soul}
\usepackage{lineno}
\usepackage{lmodern}
\usepackage[T1]{fontenc}
\usepackage{tikz-cd}
\usepackage{graphicx}
\usepackage{ragged2e}
\usepackage{relsize}
\usepackage[all]{xy}
\usepackage[mathscr]{euscript}

\usepackage{verbatim} 

\providecommand{\fibbun}[2]{\pi_{#2\! #1}}

\def\keywords#1{\vspace{10pt}
     \begin{indented}
     \item[]\rm Keywords: #1\par
     \end{indented}}

\def\AMS#1{\vspace{10pt}
     \begin{indented}
     \item[]\rm AMS classification scheme numbers: #1\par
     \end{indented}}

\begin{document}

\title{Geometric analysis for the Pontryagin action and boundary terms}

\author{Jasel Berra--Montiel$^{1}$, Iñaki de Santos$^2$ and Alberto Molgado$^{1}$}

\address{$^{1}$ Facultad de Ciencias, Universidad Autónoma de San Luis 
Potosí \\
Campus Pedregal, 
Av.~Parque Chapultepec 1610,
Col.~Privadas del Pedregal,
San Luis Potosi, SLP, 78217, Mexico}
\address{$^{2}$ Instituto de Física, Universidad Autónoma de San Luis 
Potosí \\
Campus Pedregal, 
Av.~Parque Chapultepec 1570,
Col.~Privadas del Pedregal,
San Luis Potosi, SLP, 78295, Mexico}

\eads{\mailto{jasel.berra@uaslp.mx}, \mailto{a377774@alumnos.uaslp.mx}, \mailto{alberto.molgado@uaslp.mx}}

\begin{abstract}

In this article, we analyze the Pontryagin model adopting different geometric-covariant approaches. In particular, we focus on the manner in which boundary conditions must be imposed on the background manifold in order to reproduce an unambiguous theory on the boundary. At a Lagrangian level, we describe the symmetries of the theory and construct the Lagrangian covariant momentum map which allows for an extension of Noether's theorems. Through the multisymplectic analysis we obtain the covariant momentum map associated with the action of the gauge group on the covariant multimomenta phase-space. By performing a space plus time decomposition by means of a foliation of the appropriate bundles, we are able to recover not only the $t$-instantaneous Lagrangian and Hamiltonian of the theory, but also the generator of the gauge transformations. In the polysymplectic framework we perform a Poisson-Hamilton analysis with the help of the De Donder-Weyl Hamiltonian and the Poisson-Gerstenhaber bracket. Remarkably, as long as we consider a background manifold with boundary, in all of these geometric formulations, we are able to recover the so-called differentiability conditions as a straightforward consequence of Noether's theorem.

\end{abstract}


\AMS{70S15, 70S10, 53D20.}

\section{Introduction}

One of the most useful approaches for the study of field theories is the Dirac-Hamiltonian analysis. This particular approach has the advantage of revealing the set of first and second-class constraints of the theory, thereby making evident the local degrees of freedom, as well as the  gauge symmetries associated with a given theory  \cite{QGS}. However, the starting point of the Dirac-Hamiltonian analysis is a foliation of the space-time manifold into space-like surfaces, setting aside, from the outset, the covariant nature of the theory. As pointed out in \cite{Forger1}, this issue may lead to non-covariant quantized systems when an appropriate method of quantization is applied to the original field theory. As an effort to overcome such matter, taking as a reference the De Donder-Weyl canonical theory \cite{WT}, an alternative geometric-covariant and finite-dimensional approach has recently been proposed within the context of multisymplectic analysis for field theories \cite{GIMMSY1,Crampin,DeLeon1}. From such a formalism, one can unequivocally identify the dynamic fields as local sections of an appropriate fibre bundle. The manner in which this is done is by extending to any field theoretic setup certain common characteristics of symplectic geometry. In particular, a generalization of the usual instantaneous momenta is implemented, giving rise to the introduction of the polymomenta which allow to define the manifold known as the covariant multimomenta phase-space manifold of the theory. Then, on this manifold we may also consider the so-called Poincaré-Cartan forms which are an essential part of the whole multisymplectic formalism, since they not only enable us to derive the equations of motion, but they also are fundamental to describe the symmetries associated with the theory. Further, through this description one can also construct the covariant momentum map, which, in turn, permits an extension of Noether's theorems within the multisymplectic formalism. As detailed in \cite{GIMMSY1,GIMMSY2,Fischer}, the covariant momentum map is of particular importance, since the complete set of first-class constraints of the system can be easily obtained by imposing adequate orthogonality conditions with infinitesimal generators of group symmetries, thus establishing an important link with the instantaneous Dirac-Hamiltonian approach.

Alternatively, one may introduce an equivalent geometric approach for the analysis of field theories on the covariant multimomenta phase-space by considering, as a subspace, the polymomenta phase-space, in which a preferred canonical $(n+1)$-form, known as the polysymplectic form allows the analysis of the field theory by means of the introduction of the so-called Poisson-Gerstenhaber bracket between a specific kind of differential forms \cite{IKCSPPS}. Particularly, the definition of such bracket allows a covariant Poisson-Hamilton analysis of the theory. Although this paper does not address constrictions at the polysymplectic level, a detailed discussion of the topic can be found in \cite{IKGD, Bonzom}, which we recommend for further reading.

The Pontryagin action is of particular interest in topological field theories, as it is closely related to the characterization of gauge fields through topological invariants~\cite{Blagojevic}. In addition, it plays a fundamental role in the analysis of physical observables and pseudoparticles, such as instantons~\cite{Chandia, Belavin, Mardones, EscalanteCS}. In fact, the integral of the Pontryagin term over a closed four-manifold is proportional to the second Chern class, with the proportionality constant corresponding to the instanton charge~\cite{instanton}. As a consequence, the Pontryagin action contributes only a phase factor to the path integral and does not influence the classical equations of motion.  Further, when the underlying manifold has a boundary, the Pontryagin action induces a Chern-Simons term at the boundary. This boundary contribution is crucial in the study of anomalies and topological phases. For instance, in the AdS/CFT correspondence, it plays a role in holographic anomaly cancellation~\cite{holography}, whereas in Condensed matter Physics, it describes topological insulators and the emergence of edge states~\cite{condensed}. Furthermore, the Pontryagin term is closely related to topological anomalies in gauge theories, including the chiral anomaly~\cite{chiral}. It also plays a significant role in axion electrodynamics, where it couples to a dynamical axion field, providing a theoretical framework for describing axionic dark matter models~\cite{axion}.  This converts the Pontryagin action into an ideal model for investigating the boundary conditions that must be applied to a spacetime manifold to ensure a well-defined theory on the boundary, based on the theory defined in the bulk \cite{Corichi}.

In light of the previous comments, our purpose here is to analyze the Pontryagin action by means of a covariant-geometric approach. In order to do so, we start from a geometric-covariant Lagrangian analysis, formulating the Pontryagin action in an adequate fibre bundle. A special emphasis is taken in the gauge symmetries of the theory, since they allow us to construct the Lagrangian covariant momentum map, which is directly related to Noether currents. We then also address the multisymplectic analysis of the theory, in which we construct the multisymplectic forms associated with the Pontryagin action that allow us to regard the action of the gauge symmetry as a canonical transformation. We further construct the covariant momentum map, building on the previous geometric framework. In both the multisymplectic and polysymplectic formalisms we emphasize the boundary conditions that are necessarily imposed on the space-time manifold in order to produce a well-defined theory on the boundary. Then, we proceed to the polysymplectic analysis by constructing the polymomenta associated with the field variables, and obtaining the equations of motion via the De Donder-Weyl Hamiltonian, as detailed in \cite{IKCSPPS, CPVD}. Finally, closely following the theory developed in \cite{GIMMSY1, GIMMSY2, Bonzom, Gotay1, DeLeon2}, we proceed with the space plus time decomposition of the theory, recovering the $t$-instantaneous Hamiltonian and the corresponding constraints of the theory as obtained by means of the Dirac-Hamilton formalism in \cite{Corichi, Escalante}.

\section{Geometric-covariant formalisms for classical field theory}\label{Geom Cov Formalisms}

In this section we discuss the geometric-covariant analysis of classical field theories as developed in \cite{GIMMSY1,Crampin,DeLeon1,GIMMSY2,IKCSPPS,Gotay1,DeLeon2,IKCSCHF,IKGPA}. First, we focus on the construction of the Lagrangian formalism, in which the Poincaré-Cartan forms are defined, leading to the definition of a symmetry group of the theory and the Lagrangian covariant momentum map, as well as to an extension of Noether's theorems within this approach. Next, we focus our attention in the multisymplectic formalism, in which the main results are the definition of the covariant momentum map and the covariant Legendre transform. Then, we introduce a polysymplectic structure in order to obtain a Poisson-Hamilton framework for the theory with the help of the De Donder-Weyl Hamiltonian and the Poisson-Gerstenhaber bracket. Lastly, we perform a space plus time decomposition of the theory, thus obtaining the $t$-instantaneous Lagrangian and Hamiltonian functions. It is important to remark that this section is not a comprehensive introduction to the theory, rather, it serves as a brief summary, providing the reader with an overview of the key definitions and main results, which are explored in detail in the referenced sources.

Before proceeding, we address a few matters on conventions. Given an $n$-dimensional space-time manifold without boundary $X$, a fiber bundle is denoted as a triple $(E,\fibbun{\,E}{X},X)$, where $E$ represents the total space, $X$ the base space, and $\fibbun{\,E}{X}:E\to X$ the projector map. The fibers of this bundle are the sets of the form $\fibbun{\,E}{X}^{-1}(p)$, for $p\in X$. As is usual in the literature, we will often refer to the fiber bundle just by its projector map. The set of sections of $\fibbun{\,E}{X}$ will be denoted, unless otherwise stated, as $\mathscr{E}_X$.  In order to avoid cumbersome notation, we will also often adopt the convention of denoting both a map and its pullback by the same symbol, in particular when no confusion arises. 

In the context of field theory, endowing $X$ with local coordinates $(x^\mu)$, the configuration space can be modeled by a fibre bundle $(Y,\fibbun{\,Y}{X},X)$, where the fibers of $Y$ are $m$-dimensional manifolds whose coordinates will be denoted by $(y^a)$. Dynamic fields can be identified as sections of the fibre bundle $\fibbun{\,Y}{X}$. Note that one of such sections, $\phi\in\mathscr{Y}_X$, can be locally identified by the coordinates $(x^\mu,\phi^a)$, where $\phi^a:=y^a\circ\phi$. Lagrangian dynamics can then be naturally studied in the first order jet bundle, $(J^1Y,\fibbun{\,J^1Y}{X},X)$. In order to define it, we first introduce the affine jet bundle $(J^1Y,\fibbun{\,J^1Y}{Y},Y)$, whose fiber over a point $y\in Y$ is given by $J^1_yY:=\left\{\gamma\in L(T_xX,T_yY)\middle|T_y\fibbun{\,Y}{X}\circ\gamma=\mathrm{Id}_{T_{x}X}\right\}$, where $L(T_xX,T_yY)$ is the set of linear maps from $T_xX$ to $T_yY$, being $x=\fibbun{\,Y}{X}(y)$. It is easily seen that $J^1_yY$ is an affine space modeled over the vector space $L(T_xX,V_yY)$, where $V_yY$ is the vertical tangent space of $Y$ at $y$, from which an element $\gamma\in J^1_yY$ can be identified by local coordinates $(x^\mu,y^a,y^a_\mu)$~\cite{Forger1}. The first order jet bundle is then constructed via the projector map $\fibbun{\,J^1Y}{X}:=\fibbun{\,Y}{X}\circ\fibbun{\,J^1Y}{Y}$. In particular, given a section $\phi\in\mathscr{Y}_X$, its associated tangent map at $x\in X$, $T_x\phi:T_xX\to T_{\phi(x)}Y$, is an element of $J^1_{\phi(x)}Y$. This allows to associate to $\phi$ a section $j^1\phi\in\mathscr{J}^1\mathscr{Y}_X$, known as the first jet prolongation of $\phi$, which can be locally described by $(x^\mu,\phi^a,\phi^a_\mu)$, where $\phi^a_\mu:=\partial_\mu\phi^a$. For a detailed treatment on fibre and jet bundles we refer the reader to reference \cite{Saunders}. 

Lastly, given the tangent bundle of $X$, $(TX,\fibbun{\,TX}{X},X)$, and its cotangent bundle, $(T^*X,\fibbun{\,T^*X}{X},X)$, we will denote their $k$-th exterior products as $(\Lambda^kTX,\fibbun{\,\Lambda^kTX}{X},X)$ and $(\Lambda^kT^*X,\fibbun{\,\Lambda^kT^*X}{X},X)$, respectively. Sections of the bundle $\fibbun{\,\Lambda^kTX}{X}$, that is, $k$-tangent multivector fields on $X$, will be denoted by $\mathfrak{X}^k(X)$, while the sections associated with the bundle $\fibbun{\,\Lambda^kT^*X}{X}$, $k$-forms on $X$, will be represented as $\Omega^{\,k}(X)$.
Having clarified this, we proceed with the geometric constructions of our interest, closely following the references \cite{Forger1,GIMMSY1, Crampin, DeLeon1,GIMMSY2, Fischer, Bonzom, Sardanashvily1}.

\subsection{Geometric-covariant Lagrangian formalism}\label{LF}

Let $(Y,\fibbun{\,Y}{X},X)$ be the covariant configuration space of a field theory. Then, we can define the action of the system as
\begin{equation}\label{Action}
    \mathcal{S}[\phi]:=\int_X(j^1\phi)^*\mathcal{L}\,,
\end{equation}
where $\mathcal{L}:J^1Y\to \Omega^{\,n}(X)$ is the Lagrangian density of the theory, which, in terms of the Lagrangian function, $L:J^1Y\to X$, is written locally as $\mathcal{L}:=L(x^\mu,y^a,y^a_\mu)\,d^{\,n}x$. An essential object to our study are the so-called Poincaré-Cartan forms, $\Theta^{(\mathcal{L})}\in\Omega^{\,n}(J^1Y)$ and $\Omega^{(\mathcal{L})}\in\Omega^{\,n+1}(J^1Y)$ which in local coordinates are identified as the $n$- and $(n+1)$-forms
\begin{subequations}\label{PCForms}
\begin{align}
    \Theta^{(\mathcal{L})}\!&:= \frac{\partial L}{\partial y^{a}_{\mu}}dy^{a}\wedge d^{\,n-1}x_{\mu}+\left(L-\frac{\partial L}{\partial y^{a}_{\mu}}y^{a}_{\mu}\right) d^{\,n}x\, ,\label{PCF}\\
    \Omega^{(\mathcal{L})}\!&:=
    dy^{a}\wedge d\left(\frac{\partial L}{\partial y^{a}_{\mu}}\right)\wedge d^{\,n-1}x_{\mu}-d\left(L-\frac{\partial L}{\partial y^{a}_{\mu}}y^{a}_{\mu}\right)\wedge d^{\,n}x\,,
\end{align}
\end{subequations}
respectively, where we are considering $d^{\,n-1}x_\mu:=\partial_\mu\lrcorner\, d^{\,n}x$. It is easily seen that the Poincaré-Cartan forms \eqref{PCForms} satisfy the relation $\Omega^{(\mathcal{L})}=-d\Theta^{(\mathcal{L})}$. One of the main reasons for the relevance of these forms is the following: Let $W\in \mathfrak{X}(J^1Y)$ be an arbitrary vector field on $J^1Y$ and let $\phi\in\mathscr{Y}_X$ be a critical point of \eqref{Action}. Then, the condition
\begin{equation}
    (j^1\phi)^*\left(W\lrcorner\, \Omega^{(\mathcal{L})}\right)=0
\end{equation}
is completely equivalent to the Euler-Lagrange field equations \cite{GIMMSY1,Gotay1}.

One of our purposes is to discuss the action of symmetries in the covariant approach; particularly that of a Lie group on the covariant configuration space. To this end, let $\mathcal{G}$ be a Lie group, whose Lie algebra will be denoted as $\mathfrak{g}$. The action of an element $\eta\in\mathcal{G}$ in $\fibbun{\,Y}{X}$ is given through a pair of maps $\eta_X:X\to X$ and $\eta_Y:Y\to Y$, which constitute a $\fibbun{\,Y}{X}$-bundle automorphism $(\eta_Y,\eta_X)$, that is, they satisfy the relation $\fibbun{\,Y}{X}\circ\eta_Y=\eta_X\circ\fibbun{\,Y}{X}$. If $\xi_\eta\in\mathfrak{g}$ is the infinitesimal generator of $\eta$, then the infinitesimal generators of $\eta_X$ and $\eta_Y$ are locally given by $\xi_{\eta}^X:=\xi^\mu(x)\partial_\mu\in\mathfrak{X}(X)$ and $\xi_{\eta}^Y:=\xi^\mu(x)\partial_\mu+\xi^a(x,y)\partial_a\in\mathfrak{X}(Y)$, respectively. The action of $\eta$ can be lifted to $J^1Y$ via the map $\eta_{J^1Y}:J^1Y\to J^1Y$ so that for every $\gamma\in J^1Y$, $\eta_{J^1Y}(\gamma)=T\eta_Y\circ\gamma\circ T\eta_X^{-1}$. The infinitesimal generator of this transformation is the first jet prolongation of the vector field $\xi_{\eta}^Y$, namely $\xi_\eta^{J^1Y}:=j^1\xi_{\eta}^Y\in\mathfrak{X}(J^1Y)$, and it may be found to have the local expression
\begin{equation}\label{J1YGen}
    \xi_\eta^{J^1Y}=\xi^\mu(x)\partial_\mu+\xi^a(x,y)\partial_a+ \left(\partial_\mu\xi^a(x,y)+\partial_b\xi^a(x,y)y^b_\mu-\partial_\mu\xi^\nu(x)y^a_\nu\right)\partial^\mu_a\,,
\end{equation}
where we introduced the notation $\partial^\mu_a:=\partial/\partial y^a_\mu$ \cite{GIMMSY1,Saunders,Sardanashvily1}. Having this in mind, we will say that the pair $\left(\eta_{J^1Y},\xi_\eta^{J^1Y}\right)$ constitutes an infinitesimal transformation on $J^1Y$.

As described in \cite{GIMMSY1, DeLeon1}, $\mathcal{G}$ is a symmetry group of the theory if for every $\eta\in\mathcal{G}$ the infinitesimal transformation  $\left(\eta_{J^1Y},\xi_\eta^{J^1Y}\right)$ is such that
\begin{equation}\label{symg}
    \mathfrak{L}_{\xi_\eta^{J^1Y}}\Theta^{(\mathcal{L})}=d\alpha^{(\mathcal{L})}_\eta\,,
\end{equation}
where $\alpha^{(\mathcal{L})}_\eta\in\Omega^{n-1}_1(J^1Y)$ \footnote{Note that $\Lambda^{k}_{r}T_e^{*}E:=\left\{\omega\in\Lambda^{k}T^{*}_{e}E\middle|V_1\lrcorner\cdots V_r\lrcorner\,\omega=0\,, \forall \,V_1,\ldots, V_r\in V_eE\right\}$ stands for the set of $(k;r)$-horizontal forms over a point of a total space manifold $E$, which is a fibre that defines the bundle $\fibbun{~\Lambda^k_rT^*E}{E}$. Sections of such a bundle are denoted by $\Omega^k_r(E)$ \cite{Bonzom}.}. If the relationship \eqref{symg} is satisfied, the pair $\left(\xi_\eta^{J^1Y},\alpha^{(\mathcal{L})}_\eta\right)$ is referred to as the Noether symmetry \cite{DeLeon1}. If, additionally, $d\alpha^{(\mathcal{L})}_\eta=0$, we will say that the symmetry is a natural one. Following \cite{Marsden}, given the dual pairing of $\mathfrak{g}$ and its dual Lie algebra, $\mathfrak{g}^*$, one can define the Lagrangian covariant momentum map, $J^{(\mathcal{L})}:J^1Y\to\mathfrak{g}^*\otimes \Lambda^{n-1}T^*J^1Y$, such that for $\xi_\eta\in\mathfrak{g}$,
\begin{equation}\label{Lcovmomap}
    J^{(\mathcal{L})}(\xi_\eta)=\xi_{\eta}^{
J^{1}Y}\lrcorner\,\Theta^{(\mathcal{L})}-\alpha ^{(\mathcal{L})}_{\eta}\, .
\end{equation}
For a critical point $\phi\in\mathscr{Y}_X$, the Lagrangian covariant momentum map \eqref{Lcovmomap} allows to define the Noether current 
\begin{equation}\label{Noethc}
    \mathcal{J}^{(\mathcal{L})}(\xi_\eta):=(j^1\phi)^*J^{(\mathcal{L})}(\xi_\eta)\,.
\end{equation}
It is straightforward to check that this quantity is a conserved current of the theory. This observation is, in fact, the first Noether theorem formulated through the geometric-covariant Lagrangian formalism \cite{GIMMSY1, Crampin, DeLeon1, Fischer, Sardanashvily1}. Moreover, if the pair $\left(\xi_\eta^{J^1Y},\alpha^{(\mathcal{L})}_\eta\right)$ constitutes a localizable symmetry \footnote{A localizable symmetry is a Noether symmetry that can be deformed to zero in the fibres of $J^1Y$ over a region of $X$. A more precise definition is given in \cite{Fischer,Bonzom, Marsden, Lee}.}, \eqref{Noethc} leads to the vanishing of the Noether charge over a Cauchy surface of $X$ (a compact oriented $(n-1)$-dimensional submanifold of $X$ without boundary), $\Sigma_t$,
\begin{equation}\label{Nullcharge}
\mathcal{Q}^{(\mathcal{L})}_{\,\Sigma_{t}}(\xi_\eta):=\int_{\Sigma_{t}}(j^{1}\phi\circ i_{t})^{*}J^{(\mathcal{L})}(\xi_\eta)=0\,,
\end{equation}
where $i_t:\Sigma_t\to X$ denotes the inclusion map. This last statement corresponds to the second Noether theorem \cite{Fischer}.

\subsection{Multisymplectic formalism}

Shifting our attention to a Hamiltonian-like formulation, given the covariant configuration space of the theory \eqref{Action}, $(Y, \fibbun{\,Y}{X}, X)$, the covariant multimomenta phase-space is defined as the bundle $(Z^\star,\fibbun{\,Z^\star}{Y},Y)$, where $Z^\star:=\Lambda^n_2T^*Y$. An element $\Xi\in Z^\star$ can be written in local coordinates as
\begin{equation}
    \Xi:=p\,d^{\,n}x+p^{\mu}_{a}dy^{a}\wedge d^{\,n-1}x_{\mu}\, ,
\end{equation}
where we have used $(x^\mu,y^a,p,p^\mu_a)$ as a coordinate system on $Z^\star$. Following \cite{IKCSPPS}, we will refer to the coordinates $p^\mu_a$ as the polymomenta variables.

In an analogous manner to the construction in the previous subsection, the vector space $Z^\star$ is endowed with a canonical $n$-form $\Theta^{(\mathcal{M})}\in\Omega^n(Z^\star)$, locally given by
\begin{equation}\label{XiPot}
    \Theta^{(\mathcal{M})}:=p\, d^{\,n}x + p^{\mu}_{a}dy^{a}\wedge d^{\,n-1}x_{\mu}\, .
\end{equation}
This form is known as the multisymplectic potential, since it allows to define the so-called multisymplectic $(n+1)$-form $\Omega^{(\mathcal{M})}:=-d\Theta^{(\mathcal{M})}$. In local coordinates,
\begin{equation}\label{OmegaMulti}
    \Omega^{(\mathcal{M})}=dy^{a}\wedge dp^{\mu}_{a}\wedge d^{\,n-1}x_{\mu}-dp\wedge d^{\,n}x\, .
\end{equation}

In order to describe the symmetries of the theory in the multisymplectic approach, given the covariant multimomenta phase-space $(Z^\star,\fibbun{\,Z^\star}{Y},Y)$ we define the bundle $(Z^\star,\fibbun{\,Z^\star}{X},X)$ simply by the composition $\fibbun{\,Z^\star}{X}:=\fibbun{\,Y}{X}\circ\fibbun{\,Z^\star}{Y}$. Let $\mathcal{G}$ be a symmetry group of the theory and let $\eta\in\mathcal{G}$. The $\fibbun{\,Y}{X}$-bundle automorphism $(\eta_Y,\eta_X)$ associated with $\eta$ induces a $\fibbun{\,Z^\star}{X}$-bundle automorphism $(\eta_{Z^\star},\eta_X)$, where $\eta_{Z^\star}:Z^\star\to Z^\star$ is defined by $\eta_{Z^\star}(\Xi):={\eta_Y^{-1}}^*(\Xi)$. Given the infinitesimal generators of $\eta_X$ and $\eta_Y$, $\xi^X_\eta$ and $\xi^Y_\eta$, respectively, the infinitesimal generator of $\eta_{Z^\star}$ is the vector field $\xi^{Z^\star}_\eta\in\mathfrak{X}(Z^\star)$ whose local expression is found to be
\begin{equation}
\begin{aligned}[b]
    \xi^{Z^\star}_\eta=&\,\,\xi^\mu(x)\,\partial_\mu+\xi^a(x,y)\,\partial_a-\big(p\,\partial_\mu\xi^\mu(x)+p^\mu_a\,\partial_\mu\xi^a(x,y)\big)\,\partial_p\\&-\big(p^\mu_b\,\partial_a\xi^b(x,y)-\,p^\nu_a\,\partial_\nu\xi^\mu(x)+p^\mu_a\,\partial_\nu\xi^\nu(x)\big)\,\partial^a_\mu\,,
\end{aligned}
\end{equation}
where $\partial_p:=\partial/\partial p$ and $\partial^a_\mu:=\partial/\partial p^\mu_a$. The flow of this vector field preserves the multisymplectic potential, i.e, $\mathfrak{L}_{\xi^{Z^\star}_\eta}\Theta^{(\mathcal{M})}=0$ \cite{GIMMSY1}. As detailed in \cite{DeLeon1}, given a Noether symmetry $\left(\xi_\eta^{J^1Y},\alpha^{(\mathcal{L})}_\eta\right)$, there is a unique vector field $\xi^\alpha_\eta\in\mathfrak{X}(Z^\star)$ such that $T\fibbun{\,Z^\star}{Y}(\xi^\alpha_\eta)=\xi^Y_\eta$ and
\begin{equation}
    \mathfrak{L}_{\xi^\alpha_\eta}\Theta^{(\mathcal{M})}=d\alpha^{(\mathcal{M})}_\eta\,,
\end{equation}
where $\alpha^{(\mathcal{M})}_\eta\in\Omega^{n-1}_1(Z^\star)$. This allows us to represent the action of $\eta$ on $Z^\star$ as the pair $\left(\eta^\alpha_{Z^\star},\xi^\alpha_\eta\right)$, where $\eta^\alpha_{Z^\star}:Z^\star\to Z^\star$ is the vector flow of $\xi^\alpha_\eta$.

As in the Lagrangian case, given the action of $\eta$ on $Z^\star$ $\left(\eta^\alpha_{Z^\star},\xi^\alpha_\eta\right)$, we can define a covariant momentum map $J^{(\mathcal{M})}:Z^\star\to\mathfrak{g}\otimes\Lambda^{n-1}T^{*}Z^{\star}$, whose action for $\xi_\eta\in\mathcal{G}$ is given by
\begin{equation}\label{covmomap}
    J^{(\mathcal{M})}(\xi_\eta)=\xi_\eta^\alpha\lrcorner\,\Theta^{(\mathcal{M})}-\alpha_\eta^{(\mathcal{M})}\,.
\end{equation}
The covariant momentum map \eqref{covmomap} will be later associated with the set of first class constraints of the theory.

Given the Lagrangian function $L$ that defines \eqref{Action} for a non-singular system, we can relate the Lagrangian and the multisymplectic approaches by means of the covariant Legendre transform, $\mathbb{F}\mathcal{L}:J^1Y\to Z^\star$, which is defined as
\begin{equation}
    \mathbb{F}\mathcal{L}(x^\mu,y^a,y^a_\mu):=\left(x^\mu,y^a, p=L-\frac{\partial L}{\partial y^a_\mu}y^a_\mu\,,\,p^\mu_a=\frac{\partial L}{\partial y^a_\mu}\right)\,.
\end{equation}
This allows us to induce information of the theory in the multisymplectic formalism, given some information on the Lagrangian formalism, and vice versa \cite{GIMMSY1, Crampin, DeLeon1}. Observe, for example, that the following important relations hold: $\Theta^{(\mathcal{L})}=\mathbb{F}\mathcal{L}^{*}\Theta^{(\mathcal{M})}$, $\Omega^{(\mathcal{L})}=\mathbb{F}\mathcal{L}^{*}\Theta^{(\mathcal{M})}$, and $\alpha^{(\mathcal{L})}_{\eta}=~\mathbb{F}\mathcal{L}^{*}\alpha^{(\mathcal{M})}_{\eta}$.

\subsection{Polysymplectic formalism}

Now, we introduce the polysymplectic formalism, paying special attention to the polysymplectic structure that can be defined in the so-called polymomenta phase-space, in order to construct a Poisson-Hamilton framework for the theory. To do so, we closely follow the developments presented in \cite{IKCSPPS, Bonzom, IKCSCHF, IKGPA, IKHEQFT}.

Let $(Z^\star,\fibbun{\,Z^\star}{Y},Y)$ be the covariant multimomenta phase-space. In order to define a Hamiltonian-like function, we can construct de quotient bundle $(P,\fibbun{\,P}{Y},Y)$, where $P:=Z^\star/\Lambda^{n}_1T^*Y$, whose elements have local coordinates $(x^\mu,y^a,p^\mu_a)$. Then, we define the polymomenta phase-space as the bundle $(P,\fibbun{\,P}{X},X)$, where $\fibbun{\,P}{X}:=\fibbun{\,Y}{X}\circ\fibbun{\,P}{Y}$ \cite{IKCSPPS, IKCSCHF,IKGPA}. A section $\varrho\in\mathscr{P}_X$ which covers $\phi=\fibbun{\,Y}{P}(\varrho)\in\mathscr{Y}_X$ can be represented by $(x^\mu, \phi^a, \pi^\mu_a)$, where $\pi^\mu_a:=p^\mu_a\circ\varrho$. Furthermore, $P$ also defines the bundle $(Z^\star,\fibbun{\,Z^\star}{P},P)$. A section $h\in\mathscr{Z}^\star_P$ is referred to as a Hamiltonian section, and $H:=-h^*p$ as its associated Hamiltonian function \cite{DeLeon1}.

A relation between $J^1Y$ and $P$ is given by the covariant Legendre map $\mathbb{F}\mathcal{L}_{\mathsmaller{\mathrm{DW}}}:~J^1Y\to P$ defined as
\begin{equation}\label{LegMap}
    \mathbb{F}\mathcal{L}_{\mathsmaller{\mathrm{DW}}}(x^\mu,y^a,y^a_\mu):=\left(x^\mu,y^a, p^\mu_a=\frac{\partial L}{\partial y^a_\mu}\right)\,.
\end{equation}
This map allows us to define a De Donder-Weyl Hamiltonian section $h_{\mathsmaller{\mathrm{DW}}}\in\mathscr{Z}^\star_P$ as one that satisfies the relation $h_{\mathsmaller{\mathrm{DW}}}  \circ\mathbb{F}_\mathsmaller{\mathrm{DW}}\mathcal{L}=\mathbb{F}\mathcal{L}$. As observed in \cite{Crampin}, the Hamiltonian function associated with $h_{\mathsmaller{\mathrm{DW}}}$ is identified as the De Donder-Weyl Hamiltonian and is explicitly given by
\begin{equation}
    H_{\mathsmaller{\mathrm{DW}}}(x^\mu,y^a,p^\mu_a):=p^\mu_a y^a_\mu-L(x^\mu,y^a,y^a_\mu)\,.
\end{equation}

In the polymomenta phase-space $\fibbun{\,P}{X}$ we can define the so-called polysymplectic structure
\begin{equation}
    \Omega^V:=\left[dp^\mu_a\wedge dy^a\wedge\varpi_\mu~\mathrm{mod}~\Omega^{n+1}_2(P)\right]\,,
\end{equation}
where $\varpi_\mu:=\partial_\mu\varpi$, being $\varpi:=dx^0\wedge\cdots\wedge dx^{n-1}$ the horizontal volume form on $P$. Let  $\overset{~p}{X}\in \mathfrak{X}^{\,p}(P)$ be a vertical $p$-multivector field, that is, $\overset{~p}{X}\lrcorner\,\omega=0$, $\forall\,\omega\in\Omega^p_1(P)$. We say that a horizontal $(n-p\,;1)$-form $\overset{~n-p}{F}\in\Omega^{n-p}_1(P)$ is a Hamiltonian $(n-p)$-form if it satisfies the relation
\begin{equation}
    \overset{~p}{X}\lrcorner\,\Omega^V=d^V\!\overset{~n-p}{F}\,,
\end{equation}
where $d^V:\Omega^p_q(P)\to\Omega^{p+1}_{q+1}(P)$ stands for the vertical exterior derivative. The set of Hamiltonian $p$-forms will be denoted as $\Omega^p_\mathrm{H}(P)$.

In order to construct a Poisson-Hamiltonian framework, we introduce the Poisson-Gerstenhaber bracket  $\{\![\,\cdot, \cdot\,]\!\}: \Omega^{n-p}_\mathrm{H}(P)\times \Omega^{n-q}_\mathrm{H}(P)\rightarrow \Omega^{n+1-(p+q)}_\mathrm{H}(P)$, defined by
\begin{equation}
    \{\![\overset{~n-p}{F},\overset{~n-q}{G}]\!\}:=(-1)^p\overset{~p}{X}_F\lrcorner\,\overset{~q}{X}_G\lrcorner\,\Omega^V\,
\end{equation}
where $\overset{~p}{X}_F$ and $\overset{~q}{X}_G$ are the vector fields associated with $\overset{~n-p}{F}$ and $\overset{~n-q}{G}$, respectively. A more in-depth description of the Poisson-Gerstenhaber bracket and its properties is found in \cite{Bonzom, IKCSCHF, IKGPA}. The relevance of this bracket is that for a section $\varrho\in\mathscr{P}_X$, the equations
\begin{subequations}\label{DDWHFE}
\begin{align}
    \partial_\mu\phi^a&=\varrho^*\{\![H_{\mathsmaller{\mathrm{DW}}},y^a\varpi_\mu]\!\}\,,\\
    \partial_\mu\pi^\mu_a&=\varrho^*\{\![H_{\mathsmaller{\mathrm{DW}}},p^\mu_a\varpi_\mu]\!\}\,,
\end{align}
\end{subequations}
known as the De Donder-Weyl-Hamilton field equations are completely equivalent to the Euler-Lagrange field equations, as long as the theory is non-singular \cite{IKCSCHF}. The treatment of singular systems is discussed in detail in \cite{IKGD,Bonzom}. 

For $\xi_\eta\in\mathfrak{g}$, and a solution of the De Donder-Weyl-Hamilton field equations \eqref{DDWHFE}, $\varrho\in\mathscr{P}_X$, we can define the quantity $J^{(\mathcal{P})}(\xi_\eta):=h_{\mathsmaller{\mathrm{DW}}}^*J^{(\mathcal{M})}(\xi_\eta)$, which leads to a conserved current, namely, 
\begin{equation}
    \mathcal{J}^{(\mathcal{P})}(\xi_\eta):=\varrho^*J^{(\mathcal{P})}(\xi_\eta)\,.
\end{equation}
Of particular importance is the case of localizable symmetries, since, in such case, for a Cauchy surface $\Sigma_t$, the Noether charge must vanish, that is
\begin{equation}
    \mathcal{Q}^{(\mathcal{P})}_{\,\Sigma_{t}}(\xi_\eta):=\int_{\Sigma_{t}}(\varrho\circ i_{t})^{*}J^{(\mathcal{P})}(\xi_\eta)=0\,.
\end{equation}

\subsection{Space plus time decomposition for classical field theory}

Let $\Sigma$ be a compact $(n-1)$-dimensional manifold such that a slicing of the spacetime manifold $X$ is given by a diffeomorphism $\Psi:\mathbb{R}\times\Sigma\to X$. We introduce $\mathscr{X}:=\{\Psi_t\,|\,t\in\mathbb{R}\}$ to denote the set of embeddings $\Psi_t:=\Psi(t,\cdot)$, which define the Cauchy surfaces $\Sigma_t:=\Psi_t(\Sigma)\subset X$. Let $\partial_t\in\mathfrak{X}(\mathbb{R}\times\Sigma)$ denote the infinitesimal generator of time translations. Then, $\zeta^X:=T\Psi(\partial_t)\in\mathfrak{X}(X)$ is the infinitesimal generator of the slicing $\Psi$, being $T\Psi$ the tangent map of $\Psi$. Considering the Cauchy surfaces $\Sigma_t$ to be given by level sets of $x^0$, $\Sigma_t$ can be given local coordinates $(x^i)$, $i=1,\ldots,n-1$ \cite{GIMMSY2, DeLeon2}.

Given $K$ an arbitrary manifold and $(K,\fibbun{\,K}{X},X)$ a fibre bundle, the slicing of $X$, $\Psi$, induces a compatible slicing of $K$ defined by the fibre bundle $(K_\Sigma,\fibbun{\,K_\Sigma}{\Sigma},\Sigma)$ and the diffeomorphism $\bar{\Psi}:\mathbb{R}\times K_\Sigma\to K$, such that the following diagram commutes:
\begin{equation}
\begin{tikzcd}
    {~~\mathbb{R}\times K_{\Sigma}} \arrow[d, xshift=0ex] \arrow[r, yshift=0ex, "\bar{\Psi}"] 
    & [1mm]{~K} \arrow[d, xshift=0.5mm]\\
    {\mathbb{R}\times \Sigma} \arrow[r, shorten=1mm, xshift=1mm, "\Psi"] & [1mm]{~X}
\end{tikzcd}
\end{equation}
Here, the vertical arrows denote projector maps \cite{GIMMSY2}. As previously, we define $\bar{\Psi}_t:=\bar{\Psi}_t(t,\cdot)$, $K_t:=\bar{\Psi}_t(K_\Sigma)\subset K$. Also, given $\partial_t\in\mathfrak{X}(\mathbb{R}\times K_\Sigma)$, the infinitesimal generator of $\bar{\Psi}$ is given by $\zeta^K:=T\bar{\Psi}(\partial_t)$. Note that $\zeta^K$ projects into $\zeta^X$ by means of $T\fibbun{\,K}{X}$. In light of this, the pair $(\Sigma_t,\zeta^X)$ and $(K_t,\zeta^K)$ is referred to as an infinitesimal and compatible slicing of $\fibbun{\,K}{X}$. On the restriction of $\fibbun{\,K}{X}$ to $\Sigma_t$, that is, $\fibbun{\,K_t}{\Sigma_t}$, we can identify the local coordinates $(x^i, k^a)$.

Let $\mathscr{K}_t$ denote the set of sections of the restricted bundle $\fibbun{\,K_t}{\Sigma_t}$. The collection
\begin{equation}
    \mathscr{K}^\Sigma:=\bigcup_{\Psi_t\in\mathscr{X}}\mathscr{K}_t
\end{equation}
defines a fibre bundle $(\mathscr{K}^\Sigma,\fibbun{\,\mathscr{K}^\Sigma}{\mathscr{X}},\mathscr{X})$, whose fibres are infinite-dimensional manifolds \cite{DeLeon2}. Following \cite{Fischer, ForgerSalles},  at $\kappa\in\mathscr{K}_t$ (an element of the fibre over $\Psi_t\in\mathscr{X}$), a local coordinate system is given by $(\kappa^a)$, with $\kappa^a:=k^a\circ \kappa $. We can define the tangent space of $\mathscr{K}_t$ at $\kappa$ as
\begin{equation}\label{tanspace}
    T_\kappa \mathscr{K}_t:=\left\{\dot{\kappa}:\Sigma_t\to V\!K_t\,\middle|\, \fibbun{\,V\!K_t}{K_t}\circ\dot{\kappa}=\kappa\right\}\,,
\end{equation}
being $\fibbun{\,V\!K_t}{K_t}$ the restriction of the vertical tangent bundle $\fibbun{\,V\!K}{K}$ to $K_t$. In adapted coordinates, an element $\dot{\kappa}\in T_\kappa\mathscr{K}_t$ can be written as
\begin{equation}
    \dot{\kappa}=\dot{\kappa}^a\frac{\partial}{\partial k^a}\,.
\end{equation}
Meanwhile, the cotangent space of $\mathscr{K}_t$ at $\kappa$, $T^*_\kappa K_t$ can be identified as the set of linear mappings from $T_\kappa \mathscr{K}$ to $\Lambda^{n-1}_1T^*K_t$ and an element $\tau\in T^*_\kappa K_t$ can be locally expressed as
\begin{equation}
    \tau=\tau_a dk^a\otimes d^{\,n-1}x_0\,.
\end{equation}
Then, the natural pairing of two elements $\dot{\kappa}\in T_\kappa\mathscr{K}$ and $\tau\in T^*_\kappa K_t$ is defined through the integration
\begin{equation}
    \langle\dot{\kappa}, \tau\rangle:=\int_{\Sigma_{t}}\dot{\kappa}\,\lrcorner\,\,\tau\, .
\end{equation}

Now, let $\alpha\in\Omega^{q+n-1}(K)$. We can define a differential form $\alpha_t\in\Omega^q(\mathscr{K}_t)$ as
\begin{equation}\label{difind}
    \alpha_t(\sigma)(V_1,\ldots,V_q):=\int_{\Sigma_t}\sigma^*\left(V_1\lrcorner\cdots\lrcorner V_q\lrcorner\,\alpha\right)\,,
\end{equation}
for $\sigma\in\mathscr{K}_t$ and $V_1,\ldots,V_q\in T_\sigma\mathscr{K}_t$. Therefore, we can induce differential forms on $\mathscr{K}_t$ from the ones defined on $K$ \cite{Vankerschaver}.

With the previous discussion in mind, let $(\Sigma_t,\zeta^X)$ and $(Y_t,\zeta^Y)$ be an infinitesimal and compatible $\mathcal{G}$-slicing of the covariant configuration space $\fibbun{\,Y}{X}$\footnote{This means that for $\xi_\eta\in\mathfrak{g}$, the jet prolongation $\zeta^{J^1Y}:=j^1\zeta^Y$ defines an infinitesimal symmetry of the theory \cite{Bonzom}}. On the restricted bundle $\fibbun{\,Y_t}{\Sigma_t}$ we can assign local coordinates $(x^i,y^a)$. The set of sections of $\fibbun{\,Y_t}{\Sigma_t}$, $\mathscr{Y}_t$, corresponds to the $t$-instantaneous configuration space of the theory, while the set $T\mathscr{Y}_t$ corresponds to the space of $t$-instantaneous velocities. Note that for $\varphi\in\mathscr{Y}_t$ there exists $\phi\in\mathscr{Y}_X$ such that $\varphi=\phi\circ i_t$, being $i_t:\Sigma_t\to X$ the inclusion map. Therefore, as detailed in \cite{GIMMSY2}, a local coordinate system on $T\mathscr{Y}_t$ is given by $(\varphi^a,\dot{\varphi}^a)$, where
\begin{equation}
    \dot{\varphi}^a:=(\mathscr{L}_{\zeta^Y}\phi)^a|_{\Sigma_t}=\left(T\phi\circ\zeta^X-\zeta^Y\circ\phi\right)^a\big|_{\Sigma_t}\,.
\end{equation}

Given the restriction of the bundle $\fibbun{\,J^1Y}{Y}$ to $Y_t$, $\fibbun{\,(J^1Y)_t}{Y_t}$, we define the jet decomposition map $\beta_{\zeta^Y}:(J^1Y)_t\to J^1(Y)_t\times VY_t$ by $\beta_{\zeta^Y}(x^i,y^a, y^a_\mu):=(x^i,y^a,y^a_i,\dot{y}^a)$, which makes evident that $(\mathscr{J}^1\mathscr{Y})_t$ and $T\mathscr{Y}_t$ are isomorphic \cite{Gotay1}. Particularly, for $j^1\phi\in\mathscr{J}^1\mathscr{Y}_X$ the jet prolongation of a field $\phi\in\mathscr{Y}_X$, $\beta_{\zeta^Y}(j^1\phi\circ i_t)=(j^1\varphi,\dot{\varphi})$. This leads to the definition of the $t$-instantaneous Lagrangian
\begin{equation}
    L_{t,\zeta^Y}(\varphi,\dot{\varphi}):=\int_{\Sigma_t}L(j^1\varphi,\dot{\varphi})\zeta^0d^{\,n-1}x_0\,
\end{equation}
where $\zeta^0:=\zeta^X\lrcorner\,dx^0$ \cite{GIMMSY2,Bonzom}. Therefore, it is natural to define the instantaneous Legendre transform  $\mathbb{F}\mathcal{L}_{t,\zeta^Y}(\varphi^a,\dot{\varphi}^a):T\mathscr{Y}_t\to T^*\mathscr{Y}_t$ as given by
\begin{equation}\label{instLT}
    \mathbb{F}\mathcal{L}_{t,\zeta^Y}(\varphi^a,\dot{\varphi}^a):=\left(\varphi^a, \pi^i_a=\frac{\partial L}{\partial \dot{\varphi ^a}}\right)\,,
\end{equation}
which, in turn allows to identify the $t$-primary constraint surface of the theory, $\mathscr{P}_{t,\zeta^Y}$, as the submanifold of $T^*\mathscr{Y}_t$ defined by the image of the instantaneous Legendre transform \eqref{instLT}.

Proceeding with the decomposition on the polymomenta phase-space, we first consider the restriction of the fibre bundle $\fibbun{\,Z^\star}{Y}$ to $Y_t$, $\fibbun{\,Z^\star_t}{Y_t}$. The set of sections of $\fibbun{\,Z_t}{\Sigma_t}$, $\mathscr{Z}_t$, is endowed with a presymplectic structure given by $\Omega_t\in\Omega^2(\mathscr{Z}_t)$, being $\Omega_t$ the $2$-form that $\Omega^{(\mathcal{M})}$ induces by means of \eqref{difind} \cite{Gotay1}. However, $T^*\mathscr{Y}_t$ does have a symplectic structure, and defining the map $R_t:\mathscr{Z}_t^\star\to T^*\mathscr{Y}_t$ by
\begin{equation}
    \left\langle R_{t}\left(\sigma\right),V \right\rangle:=\int_{\Sigma_{t}}\varphi^{*}\left(V\lrcorner\,\sigma\right)\, ,
\end{equation}
where $\sigma\in\mathscr{Z}^\star_t$ is such that $\varphi=\fibbun{\,Y}{Z^\star}\circ\sigma$ and $V\in T_\varphi \mathscr{Y}_t$, and noting that $\mathrm{ker}R_{t}=\left\{ \sigma\in\mathscr{Z}^{\star}_{t}\,\middle|\, p^{0}_{a}\circ\sigma=0\right\}$, one can see that the quotient map $\mathscr{Z}^\star_t/\mathrm{ker}\,TR_t\to T^*\mathscr{Y}_t$ is a symplectomorphism \cite{GIMMSY2, Gotay1}. Furthermore, the subset $\mathscr{N}_t:=\left\{\sigma\in\mathscr{Z}^\star\,\middle|\,\sigma=\mathbb{F}\mathcal{L}\circ j^1\phi\circ i_t\right\}$ projects onto $\mathscr{P}_{t,\zeta^Y}$ by means of $R_t$ \cite{Gotay1}.

Now, in order to study the action of $\mathcal{G}$ on $\mathscr{Z}^\star_t$, we introduce the energy-momentum map induced by \eqref{covmomap}, $E_t:\mathscr{Z}^\star\to\mathfrak{g}^*$, defined by
\begin{equation}
    \langle E_{t}\left(\sigma\right),\xi_{\eta}\rangle   :=\int_{\Sigma_{t}}\sigma^{*}J^{(\mathcal{M})}\left(\xi_{\eta}\right)\,,
\end{equation}
where $\sigma\in\mathscr{Z}^\star$ and $\xi_\eta\in\mathfrak{g}$. We also introduce the subset $\mathcal{G}_t:=\left\{\eta\in\mathcal{G}\,\middle|\,\eta_X(\Sigma_t)=\Sigma_t\right\}$ as the set which acts on $\Sigma_t$ by diffeomorphisms, and we will denote its Lie algebra as $\mathfrak{g}_t$. The action of $\mathcal{G}_t$ on $\mathscr{Z}^\star_t$ preserves the presymplectic structure $\Omega_t$ and, therefore, the momentum map associated with this action can be defined as $\mathcal{J}_t:=E_t|_{\mathfrak{g}_t}:\mathscr{Z}^\star_t\to\mathfrak{g}^*$ \cite{GIMMSY2}. Also, due to the symplectic structure on $\mathscr{Z}^\star_t/\mathrm{ker}\,TR_t$, the momentum map $\mathscr{J}_t:T^*\mathscr{Y}_t\to\mathfrak{g}^*_t$ corresponding to the action of $\mathcal{G}_t$ on $T^*\mathscr{Y}_t$ is given by
\begin{equation}\label{momampJ}
    \left\langle \mathscr{J}_{t}(\varphi,\pi),\xi_{\eta}\right\rangle:=\left\langle \mathcal{J}_{t}\left(\sigma\right),\xi_{\eta}\right\rangle\, ,
\end{equation}
where $\xi_\eta\in\mathfrak{g}^*$, $(\varphi,\pi)\in T^*\mathscr{Y}_t$ and $\sigma\in R_t^{-1}\{(\varphi,\pi)\}\subset\mathscr{Z}^\star_t$ \cite{GIMMSY2}.

As discussed in $\cite{GIMMSY2,Gotay1, DeLeon2}$, the $\alpha^{(\mathcal{M})}$-lift of $\zeta^Y$ to $Z^\star$, $\zeta^{Z^\star}\in\mathfrak{X}(Z^\star)$, acts on $Z^\star$ by means of covariant transformations. As a consequence, there exists a well-defined function $H_{t,\zeta^Y}:\mathscr{P}_{t,\zeta^Y}\to\mathbb{R}$, into which  $\zeta^{Z^\star}$ projects. Explicitly,
\begin{equation}\label{tinstH}
    H_{t,\zeta^Y}(\varphi,\pi):=-\int_{\Sigma_t}\sigma^*J^{(\mathcal{M})}(\zeta^{Z^\star})\,
\end{equation}
which is precisely the $t$-instantaneous Hamiltonian of the theory.

Finally, it is important to mention that for a classical field theory with localizable symmetries, the set of first-class constraints is given by the vanishing of the projected momentum map \eqref{momampJ} \cite{GIMMSY1, GIMMSY2, Fischer}, that is,
\begin{equation}
    \mathscr{J}_{t}^{-1}(0):=\left\{(\varphi,\pi)\in T^{*}\mathscr{Y}_{t}\,\middle|\, \langle \mathscr{J}_{t}(\varphi,\pi),\xi_{\eta}\rangle=0, \forall\,\xi_{\eta}\in\mathfrak{g}_{t}\right\}\, .
\end{equation}

\section{Geometric-covariant analysis of the Pontryagin action}\label{Analysis of BLmodel}

In this section, we build upon the geometric-covariant formalism introduced earlier by analyzing a classical field theory within this framework. Specifically, we examine the Pontryagin model, beginning with a brief mathematical description of the theory and its associated symmetries. By doing so, we obtain a clearer understanding of the theory at the classical level. In the spirit of \cite{Corichi}, we will place particular emphasis on the conditions under which the theory leads to well-defined boundary conditions, a critical aspect for ensuring the model's consistency when dealing with spacetime manifolds with boundary.

\subsection{ Lagrangian analysis}\label{Lagsec}

Let $\mathcal{G}$ be a compact simple Lie group, which will be the gauge group of the theory. Then, the Pontryagin action is defined by 
\begin{equation}\label{Paction}
\mathcal{S}_\mathrm{\,P}[A^a]:=\int_{X}F^a\wedge F_a ,
\end{equation}
where $X$ is a four-dimensional space-time manifold without boundary, $F^a:=dA^a+\frac{1}{2}[A\wedge A]^a$ is the curvature two-form generated by the $\mathfrak{g}$-valued connection 1-form $A^a\in \Omega^{1}(X)$, and $[\cdot\wedge\cdot]:\Omega^{\,k}(X)\times\Omega^{\,l}(X)\to\Omega^{\,k+l}(X)$ stands for the graded commutator between $\mathfrak{g}$-valued forms, whose explicit action for a pair of such forms, $(\omega,\eta)\in\Omega^{\,k}(X)\times\Omega^{\,l}(X)$, is given by $[\omega\wedge\eta]:=\omega\wedge\eta+(-1)^{kl}\,\eta\wedge\omega$.

The action of the gauge group $\mathcal{G}$ can be seen trough the transformation associated with it. In the following manner, for a $\mathfrak{g}$-valued function, $\theta^a\in\Omega^{\,0}(X)$, the symmetry of the theory is given by
\begin{equation}\label{transf}
    A^a\rightarrow A^a_\theta:=A^a+d_A\theta^a\,.
\end{equation}
Here, $d_A:\Omega^{\,k}(X)\to\Omega^{\,k+1}(X)$ is the covariant exterior derivative given by $d_A:=d+[A\wedge\cdot]$. Locally, the transformation \eqref{transf} reads
\begin{equation}
    A^a_\mu\rightarrow {A_\theta}^a_\mu=A^a_\mu+D_\mu\theta^a\,,
\end{equation}
where $D_\mu\theta^a:=\partial_\mu\theta^a+{f^a}_{bc}A^b_\mu\theta^c$, and ${f^a}_{bc}$ are the structure constants of the Lie algebra $\mathfrak{g}$.

It is readily observed that the configuration space of this theory can be identified as the fibre-bundle $\big(Y,\fibbun{\,Y}{X}, X\big)$, where $Y:=\Lambda^{1}\,T^*M$, and $A^a\in\mathscr{Y}_{X}$ is a section of $\fibbun{\,Y}{X}$. Given a local coordinate system on $X$, $(x^\mu)$, we can define an adapted coordinate system on $Y$ as $(x^\mu,a^a_\nu)$. Note that, since the dimensions of $X$ and $Y$ are the same, there is no confusion in labeling their 
respective coordinates as we did. Therefore, a section $\phi\in\mathscr{Y}_{X}$ is locally represented by $(x^\mu,A^a_\nu)$, where $A^a_\nu := a^a_\nu \circ \phi$. Moreover, being $(J^{1}Y, \fibbun{\,J^{1}Y}{Y},Y)$ the affine jet bundle over $Y$, we can define a local coordinate system on $J^{1}Y$ by $(x^\mu,a^a_\nu, a^a_{\mu\nu})$ and, in doing so, the first jet prolongation of a field $\phi$, $j^{1}\phi\in \mathscr{J}^{1}\mathscr{Y}_{X}$, can be identified by means of the local coordinate system $(x^\mu,A^a_\nu,\partial_\mu A^a_\nu)$.

Bearing this construction in mind, the action \eqref{Paction} can be rewritten as
\begin{equation}
    \mathcal{S}_\mathrm{\,P}[A^a]=\int_X d^{\,4}x\, \left(j^1\phi\right)^*\!L_{\mathrm{P}}\, ,
\end{equation}
with the Lagrangian function $L_{\mathrm{P}}:J^1Y\rightarrow \mathbb{R}$  defined by 
\begin{equation}\label{PLag}
    L_{\mathrm{P}}:=\frac{1}{4}\epsilon^{\mu\nu\rho\sigma}F^a_{\mu\nu}F_{a\rho\sigma}\, , 
\end{equation}
where $F^a_{\mu\nu}:=a^a_{\mu\nu}-a^a_{\nu\mu}+{f^a}_{bc}a^b_\mu a^c_\nu$ are the components of the curvature two-form $F^a$.

With the aid of the Lagrangian \eqref{PLag}, we can define the objects that contain the whole dynamical information of the theory, namely the Poincaré-Cartan forms \eqref{PCForms} that, in this case, are explicitly found to be
\begin{subequations}
\begin{align}
\Theta^{(\mathcal{L})}_\mathrm{P}&=\epsilon^{\mu\nu\rho\sigma}F_{a\rho\sigma}\left[da^a_\nu\wedge d^{\, 3}x_\mu+\frac{1}{2}\left({f^a}_{bc}a^b_\mu a^c_\nu-\frac{1}{2}F^a_{\mu\nu}\right)d^{\,4}x\right]\label{SigmaPL}\, ,\\
\Omega^{(\mathcal{L})}_\mathrm{P}&=-\epsilon^{\mu\nu\rho\sigma}\Bigl\{dF_{a\rho\sigma}\wedge\Bigl[da^a_\nu\wedge d^{\,3}x_\mu+\frac{1}{2}\left({f^a}_{bc}a^b_\mu a^c_\nu-F^a_{\mu\nu}\right)d^{\,4}x\Bigr]\notag\\
&~~~~~~~~~~~~~~+F_{a\rho\sigma}{f^a}_{bc}a^b_\mu da^c_\nu\wedge d^{\,4}x\Bigr\}\label{OmegaPL}\, .
\end{align}
\end{subequations}

In order to find the equations of motion, let us consider an arbitrary vector field  $W\in \mathfrak{X}\left(J^{1}Y\right)$, locally given by 
\begin{equation}
W:=W^\mu\frac{\partial}{\partial x^\mu}+W^a_\nu\frac{\partial}{\partial a^a_\nu}+W^a_{\mu\nu}\frac{\partial}{\partial a^a_{\mu\nu}}\, .
\end{equation}
Contracting this vector field with the multisymplectic form \eqref{OmegaPL} we straightforwardly obtain
\begin{equation*}
\begin{aligned}
W\!\lrcorner\,\,\Omega^{(\mathcal{L})}_\mathrm{P}=-\epsilon^{\mu\nu\rho\sigma}&\bigg\{2\big(W_{a\rho\sigma}+W^c_\sigma f_{abc}a^b_\rho\big)\Bigl[da^a_\nu\wedge d^{\,3}x_\mu+\frac{1}{2}\big({f^a}_{bc}a^b_\mu a^c_\nu-F^a_{\mu\nu}\big)d^{\,4}x\Bigr] \\
&-dF_{a\rho\sigma}\wedge\Bigl[W^a_\nu d^{\,3}x_\mu-W^\gamma da^a_\nu\wedge d^{\,2}x_{\mu\gamma}+\frac{1}{2}W^\gamma\big({f^a}_{bc}a^b_\mu a^c_\nu-F^a_{\mu\nu}\big)d^{\,3}x_\gamma\Bigr] \\
&+F_{a\rho\sigma}{f^a}_{bc}a^b_\mu\big(W^c_\nu d^{\,4}x-W^\gamma da^c_\nu\wedge d^{\,3}x_\gamma\big)\bigg\}\,.
\end{aligned}
\end{equation*}
Then, for a critical point of the theory, $\phi\in\mathscr{Y}_X$, the condition $\left(j^{1}\phi\right)^{*}\left(W\lrcorner\,\,\Omega^{(\mathcal{L})}_{\mathrm{P}}\right)=0$ gives rise to the relations
\begin{equation}\label{bianchi}
\epsilon^{\mu\nu\rho\sigma}D_\nu F^a_{\rho\sigma}=0\,,
\end{equation}
where we have used that, by definition, $\left(j^{1}\phi\right)^{*}a^a_\mu=A^a_\mu$ and $\left(j^{1}\phi\right)^{*}a^a_{\mu\nu}=\partial_\mu A^a_\nu$. Relations \eqref{bianchi} correspond to the Bianchi identities, meaning that the field equations of motion are trivially satisfied.

Let us then turn our attention to the gauge symmetries associated with the Pontryagin theory given by the symmetry group $\mathcal{G}$. Let $\xi_\theta\in\mathfrak{g}$. We define $\xi^Y_\theta\in\mathfrak{X}(Y)$ as the infinitesimal generator of the transformation \eqref{transf}. In local coordinates, it has the form
\begin{equation}\label{GeneratorY}
\xi^Y_\theta=D_\nu\theta^a\frac{\partial}{\partial a^a_\nu}\, .
\end{equation}
Then, the first jet prolongation of the infinitesimal generator of the gauge transformation, $\xi^{J^1Y}_\theta$, is obtained by means of \eqref{J1YGen} as
\begin{equation}
\xi^{J^1Y}_\theta:=j^1\xi^Y_\theta=D_\mu\theta^a\frac{\partial}{\partial a^a_\mu}+\partial_\mu D_\nu\theta^a\frac{\partial}{\partial a^a_{\mu\nu}}\, .
\end{equation}

By a direct calculation, one can verify that the action of the gauge theory completely preserves the Poincaré-Cartan form \eqref{SigmaPL}, that is, 
\begin{equation}
\mathfrak{L}_{\xi^{J^1Y}_\theta}\Theta^{(\mathcal{L})}_\mathrm{P}=0\, ,
\end{equation}
which shows that we are dealing with a natural symmetry of the theory. Since we are dealing with a localizable symmetry, the action of $\mathcal{G}$ on $J^1Y$ has an associated Lagrangian covariant momentum map, $J^{(\mathcal{L})}:J^{1}Y\rightarrow \mathfrak{g}^{*}\otimes \Lambda^{3}T^{\,*}J^{1}Y$, whose local representation for an arbitrary $\xi_\theta\in\mathfrak{g}$, $J^{(\mathcal{L})}(\xi_\theta)\in\Omega^{\,3}(J^{1}Y)$, is given by
\begin{equation}
\label{lagmomentmap}
J^{(\mathcal{L})}(\xi_\theta)=\epsilon^{\mu\nu\rho\sigma}F_{a\rho\sigma}D_\nu\theta^a d^{\, 3}x_\mu\, .
\end{equation}
The relevance of this expression is due to its role in obtaining the associated Noether current of the theory, $\mathcal{J}^{(\mathcal{L})}\!\left(\xi_\theta\right)$, obtained simply by pulling-back this covariant momentum map with a solution to the Euler-Lagrange equations of the system, that in our case is any field $\phi\in \mathscr{Y}_X$. Note that, since the symmetry is localizable, the integration of the Noether current over a Cauchy surface $\Sigma_{t}$ must vanish, as considered in \eqref{Nullcharge} above. 

So far, we have considered that $X$ is a space-time manifold without boundary. However, this issue is only manifest in the last observation, since the vanishing of the Noether current requires the absence of boundary. If we now consider $X$ to be a manifold with boundary and impose a vanishing integration of the Noether current $\mathcal{J}^{(\mathcal{L})}\!\left(\xi_\theta\right)$, we obtain as a condition that the pullback of the two-form $F^a$ to the boundary $\partial\Sigma_{t}$ must also vanish, where $\Sigma_t$ is now a $3$-dimensional submanifold of $X$ with boundary, resulting from a foliation of $X$. Therefore, in a natural manner we recover the so-called differentiability condition that was imposed in \cite{Corichi} in order to guarantee that the total Hamiltonian is a differentiable function.\footnote{See also~\cite{TeitelboimB} for a discussion on the imposition of differentiability conditions for a generic field theory with boundary.} Indeed, within our geometric formulation, these differentiability conditions are simply obtained by considering Noether's second theorem.  In addition, this result will be echoed in any of the geometric formulations considered in the following.

\subsection{Multisymplectic analysis}

We now turn our attention to the multisymplectic  analysis of the Pontryagin theory. To this end, we start by considering the covariant multimomenta phase-space of the theory, $(Z^{\star}, \fibbun{\,Z^{\star}}{Y}, Y)$. Given a coordinate system $(x^\mu,a^a_\nu)$ on $Y$, we introduce $(x^\mu,a^a_\nu,p,p_a^{\,\mu\nu})$ as a coordinate system on $Z^\star$. With this, we can write the local expressions for the canonical and multisymplectic forms, \eqref{XiPot} and \eqref{OmegaMulti}, for the Pontryagin model:
\begin{subequations}
\begin{align}
\Theta^{(\mathcal{M})}_\mathrm{P}&:=p\,d^{\, 4}x+p_a^{\,\mu\nu}da^a_\nu\wedge d^{\, 3}x_\mu\, ,\\
\Omega^{(\mathcal{M})}_\mathrm{P}&:=da^a_\nu\wedge dp_a^{\,\mu\nu}\wedge d^{\, 3}x_\mu-dp\wedge d^{\, 4}x\, .
\end{align}
\end{subequations}

Given $\xi_\theta\in\mathfrak{g}$, let us consider the $\alpha^{(\mathcal{M})}$-lift of the vector field $\xi^Y_\theta$, $\xi^\alpha_\theta\in\mathfrak{X}(Z^\star)$, as the infinitesimal generator of the gauge transformation \eqref{transf} on the covariant multimomenta phase-space. We can see that this vector field is explicitly given by
\begin{equation}
\label{inftgenonZ}
\xi^\alpha_\theta=D_\mu\theta^a\frac{\partial}{\partial a^a_\mu}-p_a^{\,\mu\nu}\partial_\mu D_\nu \theta^a\frac{\partial}{\partial p}-{f_{ab}}^c p_c^{\mu\nu}\theta^b\frac{\partial}{\partial p_a^{\mu\nu}}\, .
\end{equation}
Then, one can see that this gauge symmetry acts on $Z^\star$ by means of canonical transformations, specifically, $\mathfrak{L}_{\xi^\alpha_\theta}\Theta^{(\mathcal{M})}_\mathrm{P}=0$. Note that in this case the $\fibbun{\,Z^{\star}}{X}$-horizontal $(3\,;1)$-form on $Z^{\star}$ associated with the lift is identically zero.

Therefore, the action of $\mathcal{G}$ on $Z^\star$ has an associated covariant momentum map, $J^{(\mathcal{M})}:Z^{\star}\rightarrow \mathfrak{g}^{*}\otimes\Lambda^{n-1}T^{*}Z^{\star}$, that has the local representation
\begin{equation}
\label{multmomentmap}
J^{(\mathcal{M})}(\xi_\theta):=\xi^\alpha_\theta\lrcorner\,\Theta^{(\mathcal{M})}_\mathrm{P}-\alpha^{(\mathcal{M})}_\theta=p_a^{\, \mu\nu}D_\nu\theta^a d^{\, 3}x_\mu\, .
\end{equation}
It is easily verified that the Lagrangian covariant momentum map \eqref{lagmomentmap} can be recovered by pulling-back the covariant momentum map \eqref{multmomentmap} to $J^1Y$ with the aid of the covariant Legendre transformation.

\subsection{Polysymplectic analysis}

Consider the quotient bundle $(P,\pi_{YP},Y)$. Then, given local coordinates $(x^\mu,a^a_\nu)$ on $Y$, we denote by $(x^\mu,a^a_\nu,p_a^{\,\mu\nu})$ a coordinate system on $P$. A section $\rho\in\mathscr{P}_{X}$ can then be locally represented by $(x^\mu,A^a_\nu,\pi_a^{\mu\nu})$. By means of the Legendre map \eqref{LegMap}, it can be seen that, for the Pontryagin model, the multimomenta associated with the field variables are given by
\begin{equation}\label{polym}
p_a^{\,\mu\nu}:=\frac{\partial L_P}{\partial a^a_{\mu\nu}}=\epsilon^{\mu\nu\rho\sigma}F_{a\rho\sigma}=\epsilon^{\mu\nu\rho\sigma}(2a_{a\rho\sigma}+f_{abc}a^b_\rho a^c_\sigma)\, .
\end{equation}
Observe that this system is not singular. Indeed, one can obtain the expression for the velocities in terms of the multimomenta as
\begin{equation}\label{velocities}
    a^a_{\mu\nu}=-\frac{1}{8}\epsilon_{\mu\nu\rho\sigma}p^{a\rho\sigma}-\frac{1}{2}{f^a}_{bc}a^b_\mu a^c_\nu\,.
\end{equation}
We can then write the De Donder-Weyl Hamiltonian of the theory:
\begin{equation}
    H\!_\mathsmaller{\mathrm{DW}}:=p^{\mu\nu}_a a^a_{\mu\nu}-L_P=-\frac{1}{16}\epsilon_{\mu\nu\rho\sigma}p^{\mu\nu}_ap^{a\rho\sigma}-\frac{1}{2}{f^a}_{bc}p^{\mu\nu}_a a^b_\mu a^c_\nu\,.
\end{equation}

With this, we are in position to calculate the De Donder-Weyl-Hamilton equations of motion. Given $\varrho\in\mathscr{P}_{X}$ a local section of $\fibbun{\,P}{X}$, and remembering that $\pi^\mu_a:=p^\mu_a\circ\varrho$, we obtain
\begin{subequations}
\label{DWH eqs}
\begin{align}
\partial_\mu A^a_\nu&=\varrho^*\{\![H_\mathsmaller{\mathrm{DW}},a^a_\nu\omega_\mu]\!\}=-\frac{1}{8}\epsilon_{\mu\nu\rho\sigma}\pi^{a\rho\sigma}-\frac{1}{2}{f^a}_{bc}A^b_\mu A^c_\nu\, ,\label{DWH eqs:a}\\
\partial_\mu \pi^{\mu\nu}_a&=\varrho^*\{\![H_\mathsmaller{\mathrm{DW}},p^{\,\mu\nu}_a\omega_\mu]\!\}=f_{abc}\pi^{b\mu\nu}A^c_\nu\, .\label{DWH eqs:b}
\end{align}
\end{subequations}
Note that \eqref{DWH eqs:a} can be obtained from the definition of the polymomenta \eqref{polym}, while \eqref{DWH eqs:b} is a direct consequence of \eqref{velocities} and \eqref{DWH eqs:a}. Therefore, we have obtained trivial expressions, valid for every field. This is in accordance with the result obtained in the Lagrangian analysis of the theory, thus making manifest the equivalence between the two approaches.

Now, given $\xi_\theta\in\mathfrak{g}$, we can obtain the local representation of the covariant momentum map in the polymomenta phase-space. Pulling-back the representation of the covariant momentum map \eqref{multmomentmap} with the De Donder-Weyl Hamiltonian section $h_{\mathsmaller{\mathrm{DW}}}\in\mathscr{Z}^{\star}_{P}$ gives as a result
\begin{equation}\label{covmp}
J^{(\mathcal{P})}\!\left(\xi_{\theta}\right)=h_{\mathsmaller{\mathrm{DW}}}^*J^{(\mathcal{M})}(\xi_\theta)=p^{\,\mu\nu}_aD_\nu \theta^a \omega_{\mu}\, .
\end{equation}
Notice that, as in the Lagrangian analysis, the integration of the conserved current $\mathcal{J}^{(\mathcal{P})}(\xi_{\theta}):=\varrho^*J^{(\mathcal{P})}(\xi_\theta)$ over a Cauchy surface $\Sigma_t$ vanishes, i.e., $\mathcal{Q}^{\left(\mathcal{P}\right)}_{\,\Sigma_{t}}\!\left(\xi_\theta\right)=0$. 

As expected by consistency with the Lagrangian analysis, it is readily seen from \eqref{covmp} that if we consider $X$ to be a manifold with boundary, and impose  the vanishing of the Noether charge, we obtain the same differentiability condition as in the subsection \ref{Lagsec}, that is, the curvature two-form $F^a$ must vanish at the boundary of a $3$-dimensional submanifold of $X$, $\Sigma_t$.

\subsection{Space plus time decomposition of the Pontryagin model}

Consider $\Sigma_t$ a Cauchy surface of $X$ characterized by $x^0=t$ for some $t\in\mathbb{R}$. We identify $\zeta^X:=\partial_0\in\mathfrak{X}^1(X)$ as the infinitesimal generator of the slicing of $X$. Then, for $\xi_\theta\in\mathfrak{g}$ we introduce $(\Sigma_t,\zeta^X)$ and $(Y_t,\zeta^Y)$ as a $\mathcal{G}$-slicing of the covariant configuration space $\pi_{XY}$, where the time direction field $\zeta^Y\in\mathfrak{X}^1(Y)$ is defined as
\begin{equation}
\label{timedirfield}
\zeta^Y:=\partial_0+\xi_\theta^Y\, ,
\end{equation}
being $\xi_\theta^Y\in\mathfrak{X}^1(Y)$ as given by \eqref{GeneratorY}. Then, $(Y_t,\pi_{\Sigma_t Y_t},\Sigma_t)$ is the restriction of $\pi_{XY}$ to $Y_t$, and can be identified by the local coordinates $(x^i,a^a_\nu)$.

Therefore, given $\mathscr{Y}_{t}$ the set of sections of $\pi_{\Sigma_t Y_t}$, the $t$-instantaneous velocity space of the theory is given by $T\,\mathscr{Y}_t$, the space tangent to $\mathscr{Y}_{t}$ as defined in \eqref{tanspace}. Considering $\phi\circ i_t\in\mathscr{Y}_t$ for some $\phi\in\mathscr{Y}_X$, where $i_t:X\rightarrow\Sigma_t$ is the inclusion map, we introduce $(A^a_\nu,\dot{A}^a_\nu)$ as a coordinate system on $T\,\mathscr{Y}_t$, where the temporal derivative of the field variables is given by
\begin{equation}
\dot{A}^a_\nu=\mathfrak{L}_{\zeta^Y}A^a_\nu=\partial_0 A^a_\nu-D_\nu\theta^a\, .
\end{equation}

To perform the space plus time decomposition at a Lagrangian level, consider the restriction of $\pi_{YJ^1Y}$ to $Y_t$, that is, $\left((J^1Y)_t,\pi_{Y_t(J^1Y)_t},Y_t\right)$. Then, the jet decomposition map $\beta_{\zeta^Y}:(J^1Y)_t\rightarrow J^1(Y_t)\times VY_t$ is locally expressed as
\begin{equation}
\beta_{\zeta^Y}(x^i,a^a_\nu,a^a_{\mu\nu}):=(x^i,a^a_\nu,a^a_{i\nu}, \dot{a}^a_\nu)\, .
\end{equation}
The instantaneous functional Lagrangian of the theory is then found to be
\begin{equation}
L^{\mathrm{P}}_{t,\zeta^Y}(A^a,\dot{A}^a):=\int_{\Sigma_t}L_{\mathrm{P}}(j^1A^a,\dot{A}^a)\zeta^0d^{\, 3}x_0=\int_{\Sigma_t}\epsilon^{0ijk}F_{ajk}(\dot{A}^a_i+D_i\theta^a-D_iA^a_0) d^{\, 3}x_0\, .
\end{equation}

By means of the instantaneous Legendre transform, we can obtain then the instantaneous momenta of the system explicitly as
\begin{equation}
\pi^\mu_a:=\frac{\partial L_{\mathrm{P}}}{\partial \dot{A}^a_\mu}=\delta^\mu_i\epsilon^{0ijk}F_{ajk}\, .
\end{equation}
From this last result, we obtain the primary constraint surface of the theory $\mathscr{P}_{\,t,\,\zeta^{Y}}\subset T^{*}\mathscr{Y}_{t}$, given by
\begin{equation}\label{primaryconstraint}
\begin{aligned}[b]
\mathscr{P}_{\,t,\,\zeta^{Y}}:=&\big\lbrace (\varphi,\pi)\in T^{*}\mathscr{Y}_{t}\,\big|\, \pi^0_a=0\, , \, \pi^i_a-\epsilon^{0ijk}F_{ajk}=0 \big\rbrace \, .
\end{aligned}
\end{equation}

Now we proceed to the decomposition at the multisymplectic level. For this purpose, we introduce the restriction of the bundle $\pi_{YZ^\star}$ to $Y_t$, $(Z^\star_t,\pi_{Y_tZ^\star_t},Y_t)$. Also, we denote by $\mathscr{Z}^{\star}_{t}$ the set of sections $\pi_{\Sigma_t Z^\star_t}=\pi_{\Sigma_t Y_t}\circ\pi_{Y_t Z^\star_t}$. We identify $\zeta^{Z^\star}\in\mathfrak{X}^1(Z^\star)$  as the $\alpha^{(\mathcal{M})}$-lift of the time direction field \eqref{timedirfield}, that is,
\begin{equation}\label{Zstar}
\zeta^{Z^\star}:=\partial_0+\xi^\alpha_\theta\, ,
\end{equation}
where $\xi^\alpha_\theta$ is the $\alpha^{(\mathcal{M})}$-lift of the vector field $\xi^Y_\theta$ as given in the expression \eqref{inftgenonZ}.

Given that the gauge symmetry group of the theory, $\mathcal{G}$, acts on $\pi_{XY}$ by vertical bundle automorphisms and, in consequence, it is well-defined on $T^{*}\mathscr{Y}_{t}$, there exists an associated momentum map, $\mathscr{J}_{t}:T^{*}\mathscr{Y}_{t}\rightarrow \mathfrak{g}^{*}$. To obtain its local representation consider $\sigma\in\mathscr{Z}_t$ a section such that
$R_t(\sigma)=(A^a,\pi_a)\in T^*\mathscr{Y}_y$. Then, considering that $\mathfrak{g}_t=\mathfrak{g}$, we can obtain the action of $\mathcal{G}$ on $T^*\mathscr{Y}_t$ as given by the relation
\begin{equation}
\label{momentummaprel}
\begin{aligned}
\big\langle \mathscr{J}_{t}(A^a,\pi_a),\xi_\theta\big\rangle:=\int_{\Sigma_{t}}\sigma^{*} J^{(\mathcal{M})}\left(\xi_\theta\right)\, ,
\end{aligned}
\end{equation}
where $\xi_\theta\in\mathfrak{g}$, and $J^{(\mathcal{M})}\left(\xi_{f}\right)$ is the covariant momentum map as derived in \eqref{multmomentmap}. Identifying $\pi_a^\mu=p^{\, 0\nu}_a\circ\sigma$, and recalling the expression for the multimomenta as obtained in \eqref{polym}, we can write
\begin{equation}\label{projectedmomentum}
\big\langle \mathscr{J}_{t}(A^a,\pi_a),\xi_\theta\big\rangle=\int_{\Sigma_t}\left(\pi^0_aD_0\theta^a-\theta^aD_i\pi^i_a\right) d^{\, 3}x_0\,.
\end{equation}
Then, introducing the parametrization
\begin{subequations}
\begin{align}
    D_0\epsilon^a_0&:=D_0\theta^a\,,\\
    \epsilon^a_i&:=\partial_i\theta^a\,,\\
    D_i\epsilon^a&:=f^a_{bc}A^b_i\theta^c\,,
\end{align} 
\end{subequations}
the projection of the momentum map gives rise to the relation
\begin{equation}
    \big\langle \mathscr{J}_{t}(A^a,\pi_a),\xi_\theta\big\rangle=\int_{\Sigma_t}\left[D_0\epsilon^a_0+\epsilon^a_i\left(\pi^i_a-\epsilon^{0ijk}F_{ajk}\right)+\epsilon^aD_i\pi^i_a\right] d^{\, 3}x_0\,,
\end{equation}
which is precisely the gauge generator obtained by means of Dirac's algorithm in \cite{Escalante}. This, in turn, allows us to determine the admissible space of Cauchy data for the
evolution equations of the system determined by 
\begin{equation}
\label{admissiblecauchy}
{\mathscr{J}}_{\,t}^{-1}(0)=\left\{ (A^a,\pi_a)\subset T^{*}\mathscr{Y}_{t}\,\middle|\, \big\langle \mathscr{J}_{t}(A^a,\pi_a),\xi_\theta\big\rangle=0 \right\}\, .
\end{equation}
Therefore we obtain the constrained surface on the instantaneous phase space $T^*\mathscr{Y}_t$ defined by the first class constraints as the zero level set of the momentum map \eqref{projectedmomentum}. Specifically,
\begin{equation}\label{projectt}
    {\mathscr{J}}_{\,t}^{-1}(0)=\left\{ (A^a,\pi_a)\subset T^{*}\mathscr{Y}_{t}\,\middle|\, \pi^0_a=0, D_i\pi^i_a=0 \right\}\, .
\end{equation}
And this, in turn, elucidates that the primary constraints \eqref{primaryconstraint} are in fact the complete set of first-class constraints of the system as characterized in the Dirac-Hamiltonian formalism, since the constraint $D_i\pi^i_a=0$ is a direct consequence of $\pi^i_a-~\!\epsilon^{0ijk}F_{ajk}=0$ and, therefore, it is not an independent constraint. Furthermore, it is evident by \eqref{projectt} that the theory does not have second-class constraints. It is then a simple exercise to note that, according to \cite{QGS}, the number of local degrees of freedom adds up to $0$, making manifest that the theory under consideration is a topological field theory.

Lastly, one can verify that the well-defined function on the primary constraint surface onto which the vector field \eqref{Zstar} projects, i.e., the $t$-instantaneous Hamiltonian of the theory \eqref{tinstH}, is a combination of the first-class constraints and, therefore, it vanishes on the primary constraint surface. Indeed, a straightforward calculation yields the $t$-instantaneous Hamiltonian
\begin{equation}
        H^{\mathrm{P}}_{\,t,\,\zeta^{Y}}(A^a,\pi_a)=\int_{\Sigma_t}\left[ \dot{A}^a_0\pi^0_a-A^a_0D_i\pi^i_a-D_i\theta^a(\pi^i_a-\epsilon^{0ijk}F_{ajk})\right]d^{\,3}x_0\,.
\end{equation}
This Hamiltonian completely corresponds with the one obtained through the standard Dirac analysis~\cite{Escalante}.

\section{Conclusions}
\label{sec:Conclu}

In this article we analyzed the Pontryagin action by means of the geometric-covariant frameworks introduced as an alternative way to describe classical field theories. At a covariant Lagrangian level, after briefly describing the Pontryagin model, we obtained the Poincaré-Cartan forms associated with it, and with those forms we derived the field equations of motion that turned out to be trivial identities.  We also note that gauge symmetry emerges as a natural symmetry of the system for which we have constructed the corresponding Lagrangian covariant momentum map that encompasses the conserved quantities associated with this symmetry in a fully geometric framework. 
 Furthermore, as discussed in~\cite{Corichi}, the differentiability condition on the curvature two-form which is required to ensure that the total Hamiltonian is a well-defined differentiable function~\cite{TeitelboimB}, was naturally satisfied within our geometric framework by virtue of Noether’s second theorem.
At a multisymplectic level, our main result was the covariant momentum map, which played an essential role while performing a space plus time decomposition. Further, in the polysymplectic framework, once the polymomenta and the De Donder-Weyl Hamiltonian of the theory were obtained, we were able to use the Poisson-Gerstenhaber bracket in order to conduct a Poisson-Hamiltonian analysis and obtain the equations of motion. As in the Lagrangian approach, the differentiability condition on the curvature two-form was obtained by means of the covariant momentum map defined on the polymomenta phase-space and the vanishing of the associated Noether charge. 

Furthermore, we performed a space plus time decomposition of the theory by introducing a foliation on our space-time manifold, which induced a foliation on the different fibre bundles that defined each of the geometrical frameworks considered here. As we have shown, through a jet decomposition at the Lagrangian level, we obtained the $t$-instantaneous Lagrangian of the theory. Also, taking advantage of the instantaneous Legendre transform, we were able to introduce the primary constraint surface of the theory. Then, by means of the covariant momentum map introduced in the multisymplectic approach we obtained, through the projection of the momentum map, the same generator of infinitesimal gauge transformations as compared to the one obtained by the standard Dirac formalism. Also in contrast with the usual method, we classified the constrictions of the theory as first-class through the zero level set of the projected momentum map. Lastly, we recovered the $t$-instantaneous Hamiltonian as a combination of the first-class constraints, as in \cite{Corichi, Escalante}, which was expected due to the topological nature of the theory.

\section*{Acknowledgments}

We thank Ángel Rodríguez-López for helpulf discussions and for drawing our attention to 
reference~\cite{TeitelboimB}.  JBM and AM would like to acknowledge support from SNII CONAHCYT (Mexico). IdS was supported by a CONAHCYT (Mexico) Postgraduate Fellowship.  JBM acknowledges financial support from Marcos Moshinsky foundation (Mexico). AM acknowledges financial support from COPOCYT (Mexico) under project 2467 HCDC/2024/SE-02/16 (Convocatoria 2024-03, Fideicomiso 23871).

\end{document}